\documentclass[letterpaper, 10pt, conference]{ieeeconf}

\IEEEoverridecommandlockouts

\usepackage{graphics}
\usepackage{epsfig}
\usepackage{times}
\usepackage{amsmath}
\usepackage{amssymb}
\usepackage{amsthm}
\usepackage{upgreek}
\usepackage{mathtools}
\usepackage{multirow}
\usepackage{color}
\usepackage{subcaption}
\usepackage{cite}
\usepackage{booktabs}
\usepackage{bm}
\usepackage{enumerate}
\usepackage{lettrine}

\usepackage{caption}
\usepackage{setspace}
\usepackage{proof}

\newtheorem{thm}{Theorem}
\newtheorem{prop}{Proposition}
\newtheorem{defn}{Definition}
\newtheorem{cor}{Corollary}
\newtheorem{lem}{Lemma}
\newtheorem{assumption}{Assumption}
\newtheorem{remark}{Remark}
\newtheorem*{problem}{Problem}

\newcommand{\norm}[1]{\lVert#1\rVert}
\newcommand{\bx}{\mathbf{x}}
\newcommand{\bu}{\mathbf{u}}
\newcommand{\bd}{\mathbf{d}}
\newcommand{\bq}{\mathbf{q}}
\newcommand{\bc}{\mathbf{c}}

\title{Safe Unicycle Navigation Among Multiple Obstacles Under Bounded Disturbances: A Learning-Augmented CLF-CBF Approach (Extended Version)}

\author{Shubham~Singh$^{*}$, Anjali~Tiwari$^{*}$, Anand~Singh, and Katsuki~Fujisawa%
\thanks{$^{*}$These authors contributed equally to this work.}%
\thanks{This is an extended version of a paper accepted at the Indian Control Conference (ICC) 2027.}%
\thanks{S.~Singh, A.~Singh, and K.~Fujisawa are with the School of Computing, Institute of Science Tokyo, Tokyo~152-8550, Japan (e-mail: \texttt{\{shubham.s.ee99, singh.a.ca69, fujisawa.k.2110\}@m.isct.ac.jp}).}%
\thanks{A.~Tiwari is with the Department of Electrical Engineering, IIT Jodhpur, Jodhpur~342030, India (e-mail: \texttt{anjali.1@iitj.ac.in}).}%
}

\begin{document}
\maketitle

\begin{abstract}
This paper presents a robust control framework for safe navigation of a unicycle robot to a goal among multiple static obstacles under an unknown, time-varying disturbance.  The controller solves a single real-time quadratic program (QP) that combines a goal-reaching term with one safety term per obstacle.  The goal-reaching term is a cross-track control Lyapunov function built from body-frame tracking errors rather than a bearing angle, which keeps its Lie derivatives bounded everywhere, including the goal. Safety is encoded using robust control barrier functions, one per obstacle, each tightened by a closed-form, state-dependent disturbance bound. When stabilization and safety conflict, a slack variable relaxes only the convergence requirement, so that the safety constraints remain strictly enforced. Whenever the QP is feasible, the robust barrier constraints guarantee forward invariance of every obstacle's safe set under all admissible disturbances. Complementing this safety guarantee, a Lyapunov analysis yields an explicit ultimate bound on the distance to the goal. A learning-augmented variant replaces the fixed objective with a reinforcement-learning reference, which accelerates convergence when multiple constraints are simultaneously active. We further derive the obstacle separation below which the gap between two obstacles becomes too narrow to pass through under the robust constraints. For such gaps, a learned virtual goal for the Lyapunov function steers the robot around the obstacles without compromising safety. Simulations, a Monte Carlo study, and a comparison against a sampled-data robust baseline support the theoretical results.
\end{abstract}

\section{Introduction}\label{sec:intro}

Autonomous mobile robots increasingly operate in safety-critical domains where collision avoidance is a strict requirement, not merely a performance metric. For instance, the robot must pursue its task while satisfying hard state constraints on obstacles and workspace boundaries~\cite{wang2017safety,ames2017cbf, xiao2022hocbf}. In practice, wheel slip, terrain irregularities, and unmodeled dynamics erode these guarantees, especially when the robot is near the safe-set boundary~\cite{jankovic2018robust,dacs2025robust, dacs2026safe}. This paper designs a controller for a unicycle robot navigating multiple static obstacles under an unknown, time-varying, norm-bounded disturbance.

Control barrier functions (CBFs)~\cite{ames2019control}, when combined with control Lyapunov functions (CLFs)~\cite{sontag1999control}, encode safety and stability as constraints that are affine in the control input, so that both can be enforced within a single QP~\cite{ames2017cbf,ferraguti2022safety}. Subsequent work has extended this framework to constraints with high relative degree~\cite{nguyen2016exponential} and to systems with parametric uncertainty~\cite{taylor2020adaptive}, and in application, to multi-robot collision avoidance~\cite{wang2017safety}, navigation in cluttered environments~\cite{jang2024safe}, and high-speed geofencing~\cite{singletary2022onboard}. Despite these advances, three complications remain, and this paper addresses them.

\textit{Feasibility.}  Under bounded inputs, the safety constraints of several simultaneously active obstacles may render the QP infeasible. To address this issue, the authors in~\cite{dai2023safe} recast the problem as a smooth scalar optimization, which avoids infeasibility by construction. The approach in~\cite{keyumarsi2023lidar} mitigates deadlocks among multiple Lyapunov-based CBFs by preventing simultaneous conflicts, and~\cite{xiao2022hocbf} introduces auxiliary feasibility CBFs, a systematic remedy that comes with additional design effort and online computation. Each of these requirements has been treated individually: \cite{dai2023safe} avoids infeasibility in the absence of unknown disturbances, and \cite{breeden2023robust} combines robustness with input bounds for a single barrier, yet their combination for multiple simultaneous obstacles remains open.

\textit{Disturbance.} The standard CBF-QP assumes exact system knowledge of the system which is rarely true for nonholonomic robots subject to wheel slip and unmodeled dynamics~\cite{jankovic2018robust,alan2022issf}. Recent literature explores robust CBFs~\cite{jankovic2018robust} that add a worst-case disturbance term, Input-to-State Safety~\cite{alan2022issf} that quantifies the resulting safe-set enlargement, and~\cite{breeden2023robust} bounds inter-sampling effects for sampled-data implementations. Robust tightening therefore poses two challenges. It consumes control authority precisely where the feasibility margin is smallest, a difficulty that~\cite{11239434} leaves open, and the extent to which it reduces the free space between two obstacles has not yet been quantified.

\textit{Learning.}  Learning has been integrated with CBF-based safety layers at several points in the control loop. Some works pair an RL policy with a robust CBF, either as a differentiable projection within a model-based RL loop~\cite{emam2022safe} or to achieve disturbance-robust tracking~\cite{qi2024safe}. Others embed safety in the learning signal itself by folding a CBF into the RL cost~\cite{marvi2021safe}. A third line of work learns components of the QP directly: the model-uncertainty terms entering the CLF and CBF constraints~\cite{choi2020reinforcement}, or the QP parameters, by differentiating through the solver~\cite{xiao2023barriernet}. In contrast, we learn only the reference that appears in the QP objective. The constraints are neither learned nor modified, and the QP is treated as part of the training environment, with no gradient passed through it. Consequently, the feasibility and safety results apply to the learned controller without re-derivation.

\textit{Contributions.}
\begin{enumerate}[(C1)]
    \item We propose a cross-track CLF built from body-frame tracking errors rather than a bearing angle, whose Lie derivatives remain bounded everywhere, including at the goal. It is combined with one robust barrier constraint per obstacle, each tightened by a closed-form, state-dependent disturbance bound with smooth activation. For the resulting QP, we derive a pointwise necessary and sufficient feasibility condition that holds for any number of obstacles.
  
    \item Whenever the QP is feasible, it proves forward invariance of every safe set and an explicit ultimate bound on the goal distance. Both are corroborated by a $50$-trial Monte Carlo study and a comparison against a sampled-data robust baseline~\cite{oruganti2023robust}.
  
    \item An RL policy supplies the nominal reference in place of the fixed minimum-effort objective, while the constraint set is left unchanged. All guarantees therefore carry over intact, and convergence accelerates when several constraints are active together. A second variant learns a virtual goal for the CLF: safety is preserved by construction, and the deadlock in front of a gap that is provably too narrow to cross is avoided empirically.
\end{enumerate}

\section{Preliminaries}\label{sec:prelim}
\subsection{Notation}
$\mathbb{R}$ and $\mathbb{R}_+$ denote the reals and
non-negative reals, and $\norm{\cdot}$ is the Euclidean norm.
$\alpha:\mathbb{R}_+\to\mathbb{R}_+$ is class $\mathcal{K}$ if
strictly increasing with $\alpha(0)=0$. For differentiable
$V:\mathbb{R}^n\to \mathbb{R}$ and $\dot{\bx}=f(\bx)+g(\bx)\bu$,
the Lie derivatives are $L_fV=\nabla V^\top f(\bx)$ and
$L_gV=\nabla V^\top g(\bx)$.

\subsection{Control Barrier Functions}

\begin{defn}[CBF~\cite{ames2019control}]
\label{defn:cbf}
For $\dot{\bx}=f(\bx)+g(\bx)\bu$, a differentiable function
$h:\mathbb{R}^n\to\mathbb{R}$ is a \emph{CBF} for the safe set $\mathcal{C}=\{\bx:h(\bx)\geq 0\}$
if there exists a class-$\mathcal{K}$ function $\alpha$ such
that
\begin{equation}
  \sup_{\bu\in\mathbb{R}^m}
  \bigl[L_fh(\bx)+L_gh(\bx)\,\bu+\alpha(h(\bx))\bigr]
  \geq 0,\quad\forall\,\bx\in\mathcal{C}.
\end{equation}
\end{defn}

\begin{thm}[CBF Forward Invariance~\cite{ames2019control}]
\label{thm:cbf}
If $h$ is a CBF for $\mathcal{C}$, then any Lipschitz controller satisfying $\dot h(\bx,\bu)+\alpha(h(\bx))\geq 0$ renders $\mathcal{C}$ forward invariant.
\end{thm}

Each obstacle $i$ is later assigned a barrier $h_i$ and safe set $\mathcal{C}_i=\{\bx:h_i(\bx)\geq 0\}$ of this form; the robot is safe overall when $\bx\in\bigcap_{i=1}^M\mathcal{C}_i$.

\section{System Model and Problem Formulation}
\label{sec:problem}

\subsection{Robot Model}
Consider a unicycle robot with state $\bx=(x,y,\theta)^\top
\in\mathbb{R}^2\times(-\pi,\pi]$, where $(x,y)$ is the planar position and $\theta$ the heading, evolving under disturbed kinematics
\begin{equation}
  \dot{x}=v\cos\theta+d_1,\quad
  \dot{y}=v\sin\theta+d_2,\quad
  \dot{\theta}=\omega,
  \label{eq:unicycle}
\end{equation}
where $\bu=[v,\omega]^\top\in\mathbb{R}^2$ is the control input and $\bd=[d_1,d_2]^\top\in\mathbb{R}^2$ an unknown,
time-varying disturbance. Compactly,
\begin{equation}
  \dot{\bx}=f(\bx)+g(\bx)\bu+B\bd,
  \label{eq:sys}
\end{equation}
with $f(\bx)=\mathbf{0}$,
$g(\bx)=\begin{bmatrix}\cos\theta&0\\\sin\theta&0\\0&1
\end{bmatrix}$, $B=\begin{bmatrix}I_2\\\mathbf{0}_{1
\times2}\end{bmatrix}$; the zero row of $B$ reflects that disturbances such as wind or wheel slip displace the robot without directly torquing its heading.

\begin{assumption}
\label{ass:dist}
The disturbance is unknown but bounded i.e., there exists $\delta>0$ such that $\norm{\bd(t)}\leq\delta$, $\forall~t\geq 0$.
\end{assumption}

\subsection{Goal and Obstacles}
\begin{figure}[t]
\centering
\includegraphics[width=0.7\linewidth]{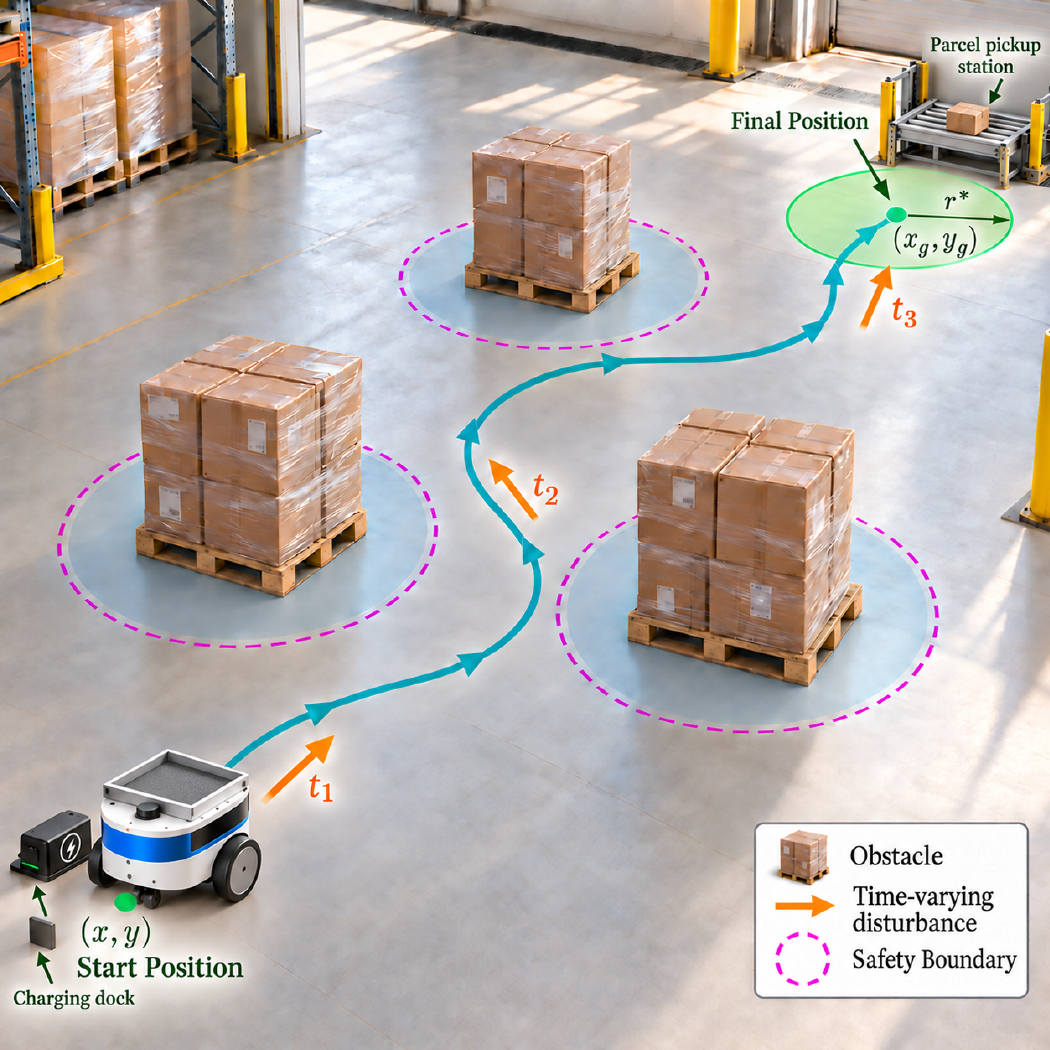}
\caption{The robot must reach the goal $(x_g,y_g)$ while avoiding the unsafe regions $\mathcal{B}_i$ under an unknown, time-varying disturbance $\bd(t)$.}
\vspace{-10pt}
\label{fig:problem}
\end{figure}

Let $\bq\triangleq(x,y)$ be the robot's position and
$\bq_g\triangleq(x_g,y_g)\in\mathbb{R}^2$ be the goal (see Fig.~\ref{fig:problem}), with goal
distance $\rho\triangleq\norm{\bq-\bq_g}$. Consider
$M\geq 1$ static circular obstacles, obstacle
$i\in\{1,\dots,M\}$ with center
$\bc_i\triangleq(c_{xi},c_{yi})$, physical radius $r_i>0$, and clearance $\rho_s>0$, giving effective radius
$\bar{r}_i\triangleq r_i+\rho_s$. The unsafe region and obstacle-free workspace are
\begin{align}
  \mathcal{B}_i &\triangleq\bigl\{(x,y):
  (x-c_{xi})^2+(y-c_{yi})^2<\bar{r}_i^2\bigr\},
  \label{eq:Bi}\\
  \mathcal{F} &\triangleq \mathbb{R}^2\setminus
  \bigcup_{i=1}^M\mathcal{B}_i.
  \label{eq:F}
\end{align}

\begin{assumption}
\label{ass:init}
$(x(0),y(0))\in\mathcal{F}$, $(x(0),y(0))\neq(x_g,y_g)$, and
$\bq_g\in\mathcal{F}$, i.e., the robot starts outside all obstacles and the goal lies outside every inflated obstacle.
\end{assumption}

\subsection{Problem Statement}
\begin{problem}
\label{prob:main}
Under Assumptions~\ref{ass:dist}--\ref{ass:init}, design a state-feedback law $\bu:\mathbb{R}^2\times(-\pi,\pi]\to
\mathbb{R}^2$ for system~\eqref{eq:sys} such that, for \emph{every} disturbance $\bd(\cdot)$ satisfying Assumption~\ref{ass:dist}, the resulting closed-loop trajectory $\bx(t)$ satisfies:
\begin{enumerate}[(P1)]
  \item \textbf{Obstacle avoidance:} the safe workspace is forward invariant, i.e., $(x(t),y(t))\in\mathcal{F}$ for
  all $t\geq 0$;
  \item \textbf{Goal reaching:} the goal distance $\rho(t)$
  is ultimately bounded by an explicit, design-dependent radius, which equals $r^*>0$, with $r^*\to 0$ as
  $\delta\to 0$, whenever the CLF slack vanishes
  asymptotically.
\end{enumerate}
\end{problem}

\section{Control Design}\label{sec:control}

\subsection{Cross-Track CLF for Goal Reaching}
\label{sec:clf}

The direct candidate $V=\rho^2$ satisfies $\partial V/
\partial\theta=0$, so $\omega$ never appears in the CLF
constraint: once the heading drifts from the goal bearing,
as it must during avoidance, the QP can only reduce $v$,
stalling the robot. A bearing-angle penalty
$\psi_\text{des}=\mathrm{atan2}(y_g-y,x_g-x)$ has gradients
scaling as $1/\rho$, undefined at the goal. We avoid both
issues via body-frame coordinates by defining the longitudinal and lateral goal-tracking errors
\begin{align}
  e_x &\triangleq (x-x_g)\cos\theta+(y-y_g)\sin\theta,
  \label{eq:ex}\\
  e_y &\triangleq -(x-x_g)\sin\theta+(y-y_g)\cos\theta,
  \label{eq:ey}
\end{align}
the projections of the goal-to-robot vector onto the heading and its perpendicular, so $e_x^2+e_y^2=\rho^2$.
Differentiating along \eqref{eq:sys},
\begin{align}
  \dot{e}_x &= v+\omega e_y+\Delta_x,
  \label{eq:exdot}\\
  \dot{e}_y &= -\omega e_x+\Delta_y,
  \label{eq:eydot}
\end{align}
where $\Delta_x=d_1\cos\theta+d_2\sin\theta$,
$\Delta_y=-d_1\sin\theta+d_2\cos\theta$ satisfy
$\Delta_x^2+\Delta_y^2=\norm{\bd}^2\leq\delta^2$, since
rotation preserves norm. The \emph{cross-track CLF} is
\begin{equation}
  V(\bx)\triangleq e_x^2+k_ye_y^2,\qquad k_y>1,
  \label{eq:V}
\end{equation}
with $k_y$ weighting lateral against longitudinal error.
Differentiating \eqref{eq:V} along
\eqref{eq:exdot}-\eqref{eq:eydot},
\begin{equation}
  \dot{V} = \underbrace{2e_xv+2(1-k_y)e_xe_y\,
  \omega}_{L_gV\bu} + \underbrace{2e_x\Delta_x+2k_ye_y
  \Delta_y}_{\Delta_V},
  \label{eq:Vdot}
\end{equation}
so
\begin{equation}
  L_g^vV = 2e_x,\qquad
  L_g^\omega V = 2(1-k_y)\,e_xe_y.
  \label{eq:LgV}
\end{equation}
Both are polynomial, bounded everywhere including at the goal, where they vanish rather than diverge; steering authority $L_g^\omega V\neq0$ is retained whenever $e_xe_y\neq0$. On the surface $e_x=0$ (goal abeam) both the terms of \eqref{eq:LgV} vanish and no input decreases $V$. To overcome this challenge, a slack is 
introduced below that absorbs the decay requirement.

\begin{lem}[$\rho$ is Dominated by $\sqrt{V}$]
\label{lem:rho_V}
Let $V$ be given by~\eqref{eq:V} with $k_y\geq 1$, and let $\rho$ denote
the distance to the goal. Then
\begin{equation}
\rho\leq\sqrt{V}\leq\sqrt{k_y}\,\rho \qquad \text{for all } \bx.
\end{equation}
\end{lem}
\begin{proof}
Since $(e_x,e_y)$ is the goal displacement expressed in the body frame
and rotations preserve length, $e_x^2+e_y^2=\rho^2$. With $k_y\geq 1$,
\[
\rho^2 = e_x^2+e_y^2 \;\leq\; e_x^2+k_ye_y^2 = V \;\leq\;
k_y\,(e_x^2+e_y^2) = k_y\rho^2,
\]
and taking square roots, which is monotone on $[0,\infty)$, yields the
claim.
\end{proof}

\begin{lem}[CLF Disturbance Bound]
\label{lem:DeltaV_bound}
For $V$ in~\eqref{eq:V} with $k_y\geq 1$ and any $\bd$
satisfying Assumption~\ref{ass:dist},
$|\Delta_V|\leq 2\delta\sqrt{k_y}\,\sqrt{V}$.
\end{lem}
\begin{proof}
Cauchy--Schwarz on $\Delta_V=2[e_x,\,k_ye_y]
[\Delta_x,\Delta_y]^\top$ gives $|\Delta_V|\leq
2\delta\sqrt{e_x^2+k_y^2e_y^2}$, and $k_y\geq1$ gives
$e_x^2+k_y^2e_y^2\leq k_y(e_x^2+k_ye_y^2)=k_yV$.
\end{proof}

The nominal decay $L_gV\bu+\eta V\leq 0$ cannot cancel the unknown $\Delta_V$ and may conflict with the safety constraints, so it is softened by a slack $\epsilon\geq 0$
\begin{equation}
  L_gV\bu+\eta V\leq\epsilon,
  \tag{CLF}
  \label{eq:CLF}
\end{equation}
penalized heavily in the QP objective.

\subsection{Robust CBFs for Multi-Obstacle Avoidance}
\label{sec:rcbf}

For obstacle $i\in\{1,\dots,M\}$, define the barrier
\begin{equation}
  h_i(\bx) \triangleq (x-c_{xi})^2+(y-c_{yi})^2-\bar{r}_i^2,
  \label{eq:hi}
\end{equation}
so $\mathcal{C}_i=\{\bx:h_i(\bx)\geq 0\}$ matches
Definition~\ref{defn:cbf}, and $h_i\geq0$ for all $i$ is
equivalent to $(x,y)\in\mathcal{F}$. Its derivative along
\eqref{eq:sys} is
\begin{equation}
  \dot{h}_i = \underbrace{2(x-c_{xi})v\cos\theta+
  2(y-c_{yi})v\sin\theta}_{L_gh_i\bu} + \Delta_i(\bx,\bd),
  \label{eq:hidot}
\end{equation}
with $\Delta_i(\bx,\bd)=2(x-c_{xi})d_1+2(y-c_{yi})d_2$.

\begin{lem}[CBF Disturbance Bound]
\label{lem:dist_bound}
For all $\bx$ with $h_i(\bx)\geq-\bar r_i^2$ and all $\bd$
satisfying Assumption~\ref{ass:dist},
\begin{equation}
  |\Delta_i(\bx,\bd)| \leq \bar{\Delta}_i(\bx) \triangleq
  2\delta\sqrt{h_i(\bx)+\bar{r}_i^2}.
  \label{eq:Delta_i_bound}
\end{equation}
\end{lem}
\begin{proof}
By \eqref{eq:hi},
$(x-c_{xi})^2+(y-c_{yi})^2=h_i+\bar{r}_i^2$. On applying the Cauchy--Schwarz we obtain \eqref{eq:Delta_i_bound}.
\end{proof}

Since $\bar\Delta_i$ grows as $\sqrt{h_i+\bar r_i^2}$, it
would over-tighten far from the obstacle; we activate the
tightening only nearby, via
\begin{equation}
  \sigma_i(h_i) \triangleq
  \begin{cases}
    1, & h_i\leq 0,\\[2pt]
    \dfrac{1}{2}\Bigl(1+\cos\dfrac{\pi h_i}{h_\text{act}}\Bigr),
    & 0<h_i<h_\text{act},\\[6pt]
    0, & h_i\geq h_\text{act},
  \end{cases}
  \label{eq:sigma_i}
\end{equation}
for a design constant $h_\text{act}>0$. The \emph{robust
CBF} (RCBF) constraint for obstacle $i$ is
\begin{equation}
  L_gh_i \bu+\gamma h_i \geq \sigma_i\bar\Delta_i(\bx),
  \tag{RCBF-$i$}
  \label{eq:RCBF}
\end{equation}
with $\gamma>0$ ($\alpha(h_i)=\gamma h_i$).

\begin{thm}[Robust Safety]
\label{thm:safety}
Let Assumptions~\ref{ass:dist}--\ref{ass:init} hold, and suppose that
the control input $\bu(t)$ satisfies the robust barrier
constraint~\eqref{eq:RCBF} for every obstacle $i\in\{1,\dots,N\}$ and
all $t\geq 0$. Then, for every disturbance satisfying
Assumption~\ref{ass:dist},
\begin{equation}
h_i(\bx(t))\geq 0 \quad \text{for all } i\in\{1,\dots,N\}
\text{ and all } t\geq 0,
\end{equation}
that is, the robot remains in the obstacle-free set,
$(x(t),y(t))\in\mathcal{F}$ for all $t\geq 0$. Hence, the safety
requirement of Problem~(P1) is satisfied.
\end{thm}
\begin{proof}
By Nagumo's theorem~\cite{blanchini1999set}, it suffices to
show $\dot h_i\ge0$ whenever $h_i=0$, since
Assumption~\ref{ass:init} gives $h_i(\bx(0))\ge0$.
By~\eqref{eq:sigma_i}, $\sigma_i=1$ at $h_i=0$, where
constraint~\eqref{eq:RCBF} reads
$L_gh_i\bu\geq\bar\Delta_i-\gamma h_i$.
Using~\eqref{eq:hidot} and Lemma~\ref{lem:dist_bound},
\begin{equation}
  \dot h_i = L_gh_i \bu+\Delta_i
  \geq -\gamma h_i+(\bar\Delta_i-|\Delta_i|)
  \geq -\gamma h_i,
\end{equation}
so $h_i{=}0\Rightarrow\dot h_i\geq0$, and $\mathcal{C}_i$ is
forward invariant for all $i$.
\end{proof}

\begin{remark}
\label{rem:sigma_continuity}
$\sigma_i$ is $C^1$ everywhere, including at the junctions
$h_i=0$ and $h_i=h_\text{act}$, where both one-sided
derivatives vanish, so the tightening
$\sigma_i(h_i)\bar\Delta_i(\bx)$ in~\eqref{eq:RCBF} is
locally Lipschitz on $\mathcal{F}$
($\bar\Delta_i$ is smooth wherever $h_i+\bar r_i^2>0$). The
safety proof uses only $\sigma_i(0)=1$; choosing
$2\delta\sqrt{h_\text{act}+\bar r_i^2}\ll\gamma
h_\text{act}$ keeps the tightening modest across the
transition band.
\end{remark}

\subsection{CLF-CBF Quadratic Program}
\label{sec:qp}

At each instant, the robot solves
\begin{subequations}\label{eq:QP}
\begin{align}
  (v^*,\omega^*,\epsilon^*) &= \arg\min_{v,\omega,\epsilon}
  \;\tfrac12(v^2+\lambda_\omega\omega^2)+p\epsilon^2,
  \label{eq:QP_obj}\\
  \text{s.t.}\quad
  &L_gV \bu+\eta V\leq\epsilon, \tag{CLF}\\
  &L_gh_i \bu+\gamma h_i\geq\sigma_i\bar\Delta_i(\bx),
  \;\; i=1,\dots,M, \tag{RCBF-$i$}\\
  &\epsilon\geq 0, \label{eq:QP_slack}\\
  &v_\text{min}(\rho)\leq v\leq v_{\max},\quad
  |\omega|\leq\omega_{\max}, \label{eq:QP_bounds}
\end{align}
\end{subequations}
where $\lambda_\omega>0$ weights angular effort, $p\gg1$
penalizes slack, and
$v_\text{min}(\rho)\triangleq v_0\min(1,\rho/
\rho_\text{stop})$ ramps to $v_0$ over
$[0,\rho_\text{stop}]$, letting the robot slow near the
goal. Here $L_gV\bu$ is given by \eqref{eq:LgV}, and
\begin{equation}
  L_gh_i\bu = \bigl[2(x-c_{xi})\cos\theta+
  2(y-c_{yi})\sin\theta,\;0\bigr]\bu.
  \label{eq:Lghi}
\end{equation}
Objective \eqref{eq:QP_obj} is strictly convex, so
\eqref{eq:QP} has a unique solution whenever feasible.

\begin{thm}[Pointwise QP Feasibility]
\label{thm:feasibility}
Fix $\bx$. By~\eqref{eq:Lghi}, $L_g^\omega h_i\equiv0$: no
RCBF constraint restricts $\omega$. For $i=1,\dots,M$ write
$a_i\triangleq L_g^vh_i$ and
$b_i\triangleq\sigma_i\bar\Delta_i-\gamma h_i$. Let
$I^+\triangleq\{i:a_i>0\}$, $I^-\triangleq\{i:a_i<0\}$ and
$I^0\triangleq\{i:a_i=0\}$, and set $L_i\triangleq b_i/a_i$
for $i\in I^+$, $U_i\triangleq b_i/a_i$ for $i\in I^-$. Then
QP~\eqref{eq:QP} is feasible at $\bx$ iff $b_i\leq0$ for all
$i\in I^0$ and
\begin{equation}
  \max\Bigl\{v_\text{min}(\rho),\max_{i\in I^+}L_i\Bigr\}
  \;\leq\;
  \min\Bigl\{v_{\max},\min_{i\in I^-}U_i\Bigr\}.
  \label{eq:feas_general}
\end{equation}
\end{thm}
\begin{proof}
Each RCBF-$i$ reads $a_iv\geq b_i$, a condition on $v$ alone
since $L_g^\omega h_i=0$: $v\geq L_i$ if $a_i>0$,
$v\leq U_i$ if $a_i<0$, and $0\geq b_i$, independent of the
input, if $a_i=0$. Joint satisfiability with
$v\in[v_\text{min}(\rho),v_{\max}]$ is exactly
\eqref{eq:feas_general} together with $b_i\leq0$ on $I^0$.
$\omega$ is free in $[-\omega_{\max},\omega_{\max}]$, and the
CLF constraint is satisfiable for any $(v,\omega)$ by taking
$\epsilon$ large; hence the condition is necessary and
sufficient.
\end{proof}

\begin{cor}[Single Tightened Constraint]
\label{cor:single-active}
If
$\norm{\bc_i-\bc_j}\geq\sqrt{h_\text{act}+\bar r_i^2}+
\sqrt{h_\text{act}+\bar r_j^2}$ for every $i\neq j$, then
$\sigma_i(\bx)>0$ for at most one $i$ at every $\bx$: at most
one constraint carries the disturbance tightening, and
$b_j=-\gamma h_j$ for every other obstacle.
\end{cor}

\begin{remark}
\label{rem:pointwise}
Theorem~\ref{thm:feasibility} is a pointwise test as it characterizes feasibility at a given state but does not by itself guarantee feasibility along the closed-loop trajectory. In particular, close to an obstacle with the heading pointing toward it ($a_i<0$, $b_i>0$), it gives $U_i<0$, which conflicts with $v\geq v_\text{min}(\rho)$. The guarantees below therefore apply to intervals where \eqref{eq:QP} is feasible, and feasibility is monitored in simulation.
\end{remark}

\begin{cor}[Closed Gap]
\label{cor:gap}
Let $D_{ij}\triangleq\norm{\bc_i-\bc_j}$ and
$\kappa\triangleq4\delta/\gamma$, and suppose $\sigma_i=\sigma_j=1$
on the part of the segment $[\bc_i,\bc_j]$ outside both inflated
obstacles. If
\begin{equation}
  D_{ij}<D^\text{pass}_{ij}\triangleq\tfrac12\Bigl(\kappa
  +\sqrt{\kappa^2+4(\bar r_i+\bar r_j)^2}\Bigr),
  \label{eq:Dpass}
\end{equation}
then RCBF-$i$ and RCBF-$j$ cannot both be satisfied at any point of that part of the segment or in a neighbourhood of it, so no trajectory along which~\eqref{eq:QP} is feasible passes between
obstacles $i$ and $j$.
\end{cor}
\begin{proof}
At the point of the segment at distance $s$ from $\bc_i$, let
$\mathbf{m}$ be the unit vector from $\bc_i$ to $\bc_j$ and
$c\triangleq\mathbf{m}^\top[\cos\theta,\sin\theta]^\top$. Then
$a_i=2sc$ and $a_j=-2(D_{ij}-s)c$, so the two constraints read
$cv\geq\beta_i$ and $-cv\geq\beta_j$, with
$\beta_i\triangleq b_i/(2s)=\delta-\frac{\gamma}{2}(s-\bar
r_i^2/s)$ and $\beta_j$ defined likewise with $D_{ij}-s$. Adding
them requires $\beta_i+\beta_j\leq0$. The sum equals
$2\delta-\frac{\gamma}{2}[D_{ij}-\bar r_i^2/s-\bar
r_j^2/(D_{ij}-s)]$, whose minimum over $s$, attained at
$s=D_{ij}\bar r_i/(\bar r_i+\bar r_j)$, is
$2\delta-\frac{\gamma}{2}[D_{ij}-(\bar r_i+\bar r_j)^2/D_{ij}]$;
this is positive iff~\eqref{eq:Dpass} holds. Since the sum is then
bounded away from zero on the segment and $|v|\leq v_{\max}$, the
two constraints remain incompatible on a neighbourhood of it by
continuity.
\end{proof}

The threshold exceeds the geometric limit $\bar r_i+\bar r_j$ by
about $\kappa/2=2\delta/\gamma$: a gap that is physically open can
be closed for the robust controller.

\section{Stability Analysis}\label{sec:stability}

\subsection{KKT Conditions}
Let $\mu\geq 0$ and $\lambda_i\geq 0$ be the KKT multipliers of the CLF and RCBF-$i$ constraints of~\eqref{eq:QP}, with input bounds~\eqref{eq:QP_bounds} momentarily inactive. The stationarity condition of the Lagrangian in $(v,\omega,\epsilon)$
gives $v^*=-\mu L_g^vV+\sum_i\lambda_i(L_gh_i)_1$,
$\omega^*=\lambda_\omega^{-1}[-\mu L_g^\omega V+
\sum_i\lambda_i(L_gh_i)_2]$, and
\begin{equation}
  \epsilon^* = \frac{\mu}{2p}.
  \label{eq:kkt_eps}
\end{equation}
Since $L_g^\omega V=2(1-k_y)e_xe_y$, a positive $\mu$
produces a nonzero $\omega^*$ whenever $e_xe_y\neq 0$,
confirming genuine steering authority without dividing by
$\rho$. When an input bound is active, e.g.\
$\omega=\pm\omega_{\max}$ during sharp avoidance, the
corresponding component saturates and the bound is absorbed
by an additional non-negative multiplier.

\subsection{Learning-Augmented Reference}
\label{sec:rl}

Theorem~\ref{thm:feasibility} characterizes feasibility but not the size of the feasible interval in~\eqref{eq:feas_general}. When two or more obstacles are active, it can collapse to a narrow band near $v=0$, which objective~\eqref{eq:QP_obj} cannot anticipate,
producing near-deadlocks with $(v^*,\omega^*)\approx(0,0)$. We address this without altering a single constraint in~\eqref{eq:QP}.

Let $\pi_\phi:\mathcal{S}\to\mathbb{R}^2$ be a policy with weights $\phi$ mapping the local observation
\begin{equation}
  s_t \triangleq \bigl(e_x,e_y,\theta,\,
  \{d_j,\beta_j,\sigma_j\}_{j\in\mathcal{N}_K(\bx)}\bigr)
  \label{eq:rl_state}
\end{equation}
to a reference $(v_\text{RL},\omega_\text{RL})=\pi_\phi(s_t)$,
where $d_j,\beta_j$ are the distance and bearing to the $j$-th of the $K$ nearest obstacles $\mathcal{N}_K(\bx)$ and $\sigma_j$ its activation~\eqref{eq:sigma_i}. The reference
enters only the QP objective: \eqref{eq:QP_obj} is replaced by
\begin{equation}
  \min_{v,\omega,\epsilon}\;
  \tfrac12 q_v(v-v_\text{RL})^2
  +\tfrac{\lambda_\omega}{2}(\omega-\omega_\text{RL})^2
  +p\epsilon^2,
  \tag{QP-obj$'$}
  \label{eq:QP_obj_RL}
\end{equation}
with $q_v>0$; constraints~\eqref{eq:CLF}--\eqref{eq:QP_bounds}
are unchanged. This problem is referred to as QP$'$.

\begin{prop}[Safety-Transparency]
\label{prop:transparent}
For any policy $\pi_\phi$ and any $t\geq0$: (a) QP$'$ is feasible whenever QP~\eqref{eq:QP} is, and Theorem~\ref{thm:feasibility} applies to QP$'$ unchanged; (b) Theorems~\ref{thm:safety} and~\ref{thm:stability} hold for the closed loop under QP$'$ without modification.
\end{prop}
\begin{proof}
Both problems share the constraint set
\eqref{eq:CLF}-\eqref{eq:QP_bounds} verbatim; changing only the objective leaves feasibility unaffected, proving (a). Theorems~\ref{thm:safety} and~\ref{thm:stability} are proved solely from constraint satisfaction, \eqref{eq:RCBF} at $h_i=0$ for safety and \eqref{eq:CLF} for the $\dot V$ bound, never referencing the objective; the argument applies verbatim to any optimal point of QP$'$, proving (b).
\end{proof}

\begin{remark}
The learned reference is thus safe by construction, not by
training: wherever the QP is feasible, even an adversarial
$\pi_\phi$ leaves every safe set forward invariant. Training
affects \emph{how quickly} the goal is approached and how
large the slack $\epsilon^*$ is, and hence the size of the
bound~\eqref{eq:UUB}, never \emph{whether} safety holds.
\end{remark}

$\pi_\phi$ is trained via Proximal Policy Optimization
(PPO)~\cite{schulman2017proximal} in the closed loop under
QP$'$, with per-step reward
\begin{equation}
  r_t \triangleq w_1\Delta\rho_t - w_2\,\epsilon_t^*
  - w_3\,\mathbf{1}\!\left[\norm{(v_t^*,\omega_t^*)}
  <v_\text{dead}\right]-w_4\norm{\bu_\text{RL}}^2,
  \label{eq:reward}
\end{equation}
rewarding progress $\Delta\rho_t=\rho(t-\Delta t)-\rho(t)$
and penalizing slack, near-zero velocity (a deadlock
signature, $v_\text{dead}\ll v_0$), and effort. Since
$(v^*,\omega^*,\epsilon^*)$ are the \emph{filtered}
quantities, $\pi_\phi$ is trained against actual
closed-loop behavior, with the QP treated as part of the
environment.

\subsection{Learned Virtual Goal}
\label{sec:vg}

The reference of Section~\ref{sec:rl} acts only on the objective
and cannot change where the CLF steers. When the direct path is
blocked, for instance by a gap closed in the sense of
Corollary~\ref{cor:gap}, the RCBFs, which act on speed alone, stop
the robot while the CLF holds its heading on the goal line: the
closed loop deadlocks, whatever the reference. We therefore let a policy
$\varphi_\phi:\mathcal{S}\to[-\tfrac{\pi}{2},\tfrac{\pi}{2}]$
rotate the goal bearing, defining the \emph{virtual goal}
\begin{equation}
  \bq_v\triangleq\bq+R\bigl(s(\rho)\,\varphi_\phi(s_t)\bigr)
  (\bq_g-\bq),\quad
  s(\rho)\triangleq\min\{1,\rho/(2r^*)\},
  \label{eq:vgoal}
\end{equation}
with $R(\cdot)$ the planar rotation, and build the errors
\eqref{eq:ex}--\eqref{eq:ey} and the CLF~\eqref{eq:V} on $\bq_v$
instead of $\bq_g$. QP~\eqref{eq:QP} is otherwise unchanged. Since
a rotation preserves length, $\norm{\bq-\bq_v}=\rho$, and
$\bq_v=\bq_g$ when $\varphi_\phi=0$.

\begin{prop}[Safety Under a Virtual Goal]
\label{prop:vg}
For any policy $\varphi_\phi$: (a) the RCBF constraints, the input
bounds and $v_\text{min}(\rho)$ in~\eqref{eq:QP} do not depend on
$\varphi_\phi$, so the feasibility condition of
Theorem~\ref{thm:feasibility} is unchanged; (b)
Theorem~\ref{thm:safety} holds for the closed loop without
modification.
\end{prop}
\begin{proof}
(a) $L_gh_i$, $h_i$, $\sigma_i$ and $\bar\Delta_i$ depend on the
state and the obstacles only, and $\norm{\bq-\bq_v}=\rho$ leaves
$v_\text{min}$ unchanged; the CLF constraint is satisfiable by the
slack for any target. (b) The proof of
Theorem~\ref{thm:safety} uses only~\eqref{eq:RCBF}.
\end{proof}

\begin{remark}
\label{rem:vg}
Unlike Proposition~\ref{prop:transparent}, no convergence
guarantee is inherited: the CLF now decreases toward a moving
target, so Theorem~\ref{thm:stability} applies only on intervals
where $\varphi_\phi=0$. Goal reaching under a virtual goal is
therefore assessed empirically in Section~\ref{sec:sim_vg}.
\end{remark}

\begin{figure*}[t]
\centering
\begin{subfigure}[t]{0.275\textwidth}
  \includegraphics[width=\linewidth]{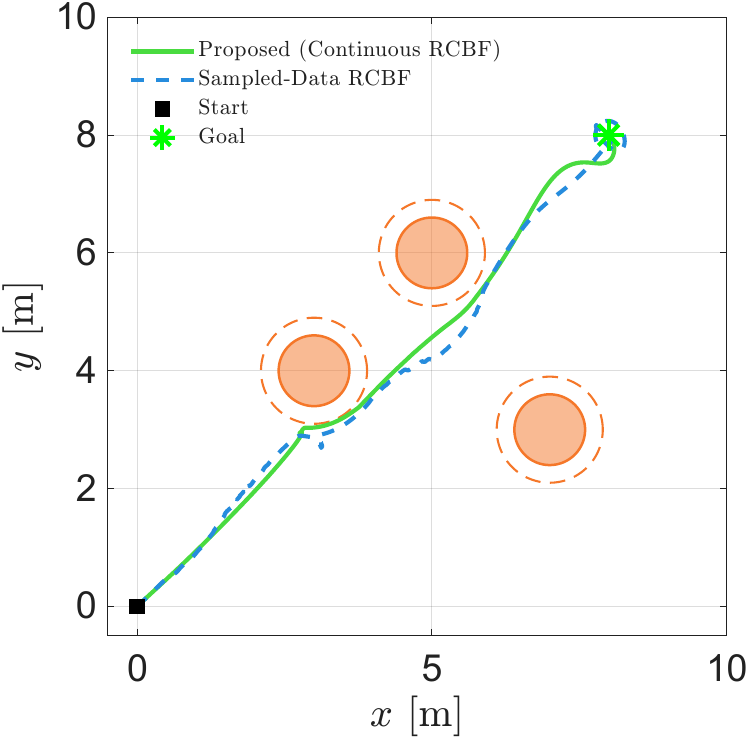}
  \caption{Robot trajectory}
  \label{fig:results_traj}
\end{subfigure}
\hfill
\begin{subfigure}[t]{0.325\textwidth}
  \includegraphics[width=\linewidth]{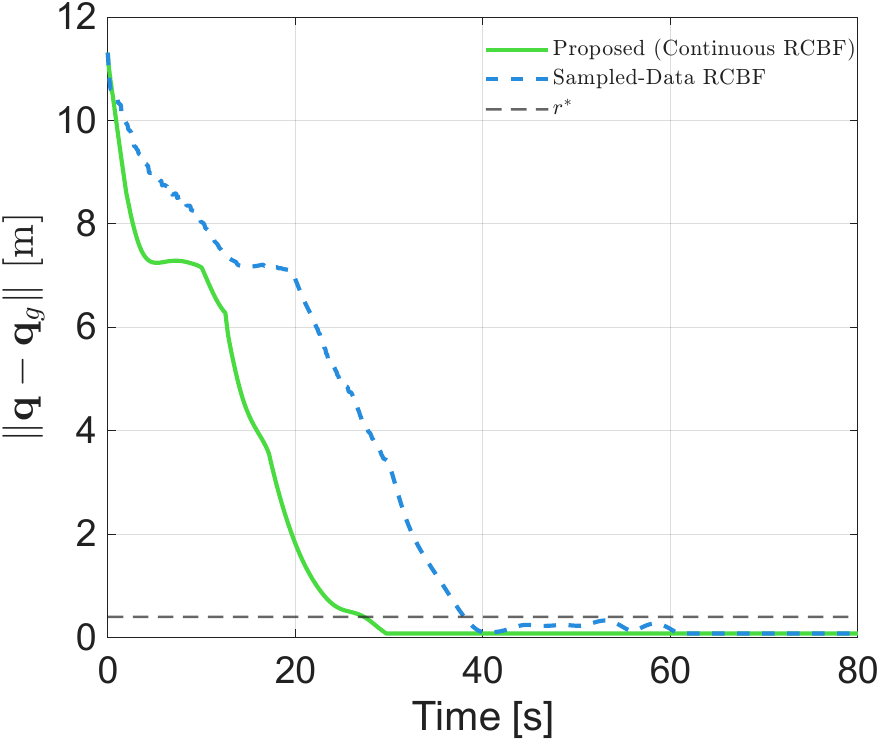}
  \caption{Goal distance $\rho(t)$}
  \label{fig:results_dist}
\end{subfigure}
\hfill
\begin{subfigure}[t]{0.32\textwidth}
  \includegraphics[width=\linewidth]{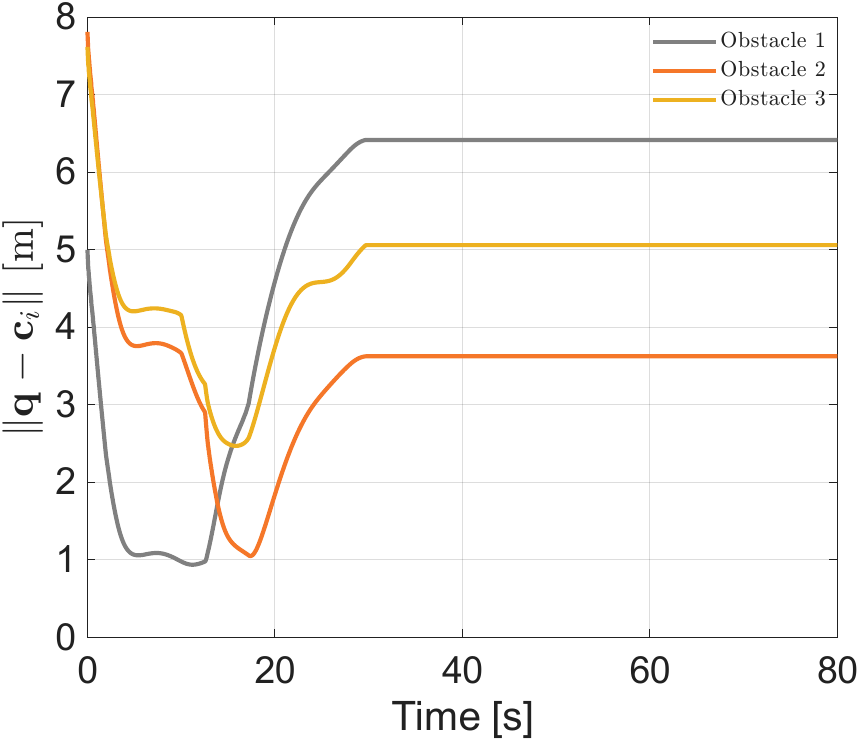}
  \caption{Distance from each obstacle}
  \label{fig:results_cbf}
\end{subfigure}
\caption{(a) trajectory bypassing all three obstacles and converging
to the goal; (b) goal distance $\rho(t)$ settling within the
radius $r^*=0.80\,\text{m}$;
(c) Euclidean distance from each obstacle, remaining above
its effective radius $\bar r_i$ throughout.}
\label{fig:results}
\end{figure*}

\begin{figure*}[t]
\centering
\begin{subfigure}[t]{0.25\linewidth}
  \includegraphics[width=\linewidth]{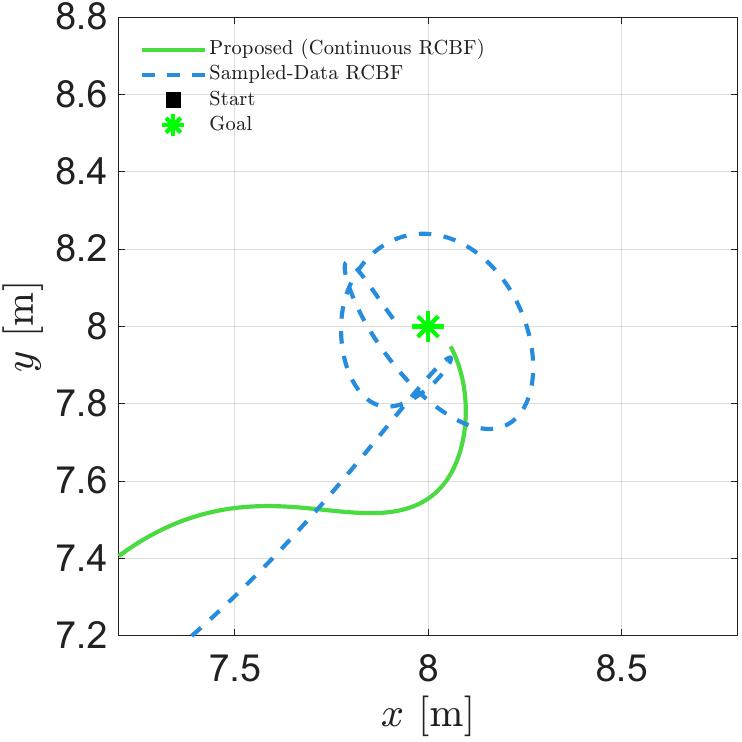}
  \caption{Near-goal zoom}
  \label{fig:traj_zoom}
\end{subfigure}
\hfill
\begin{subfigure}[t]{0.25\linewidth}
  \includegraphics[width=\linewidth]{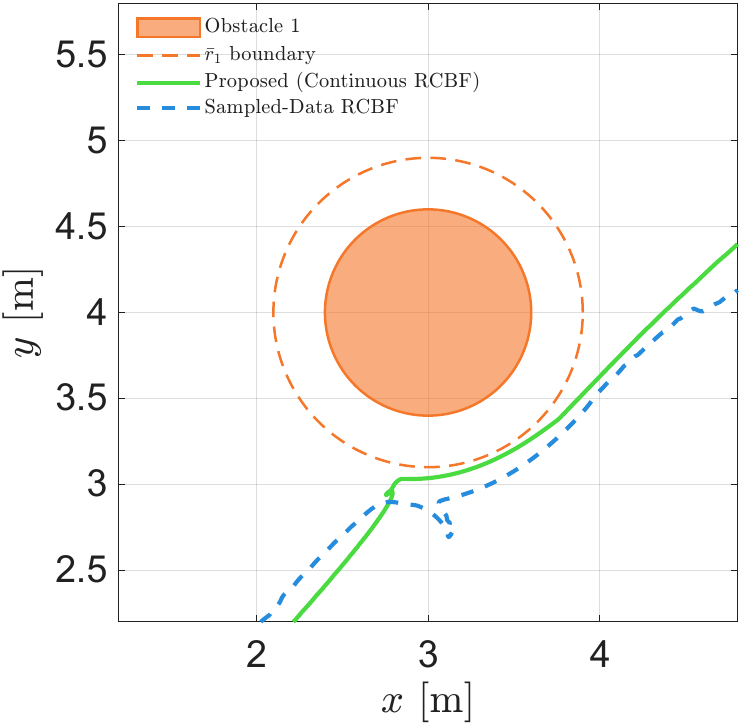}
  \caption{Obstacle-avoidance zoom}
  \label{fig:obs_zoom}
\end{subfigure}
\hfill
\begin{subfigure}[t]{0.29\linewidth}
  \includegraphics[width=\linewidth]{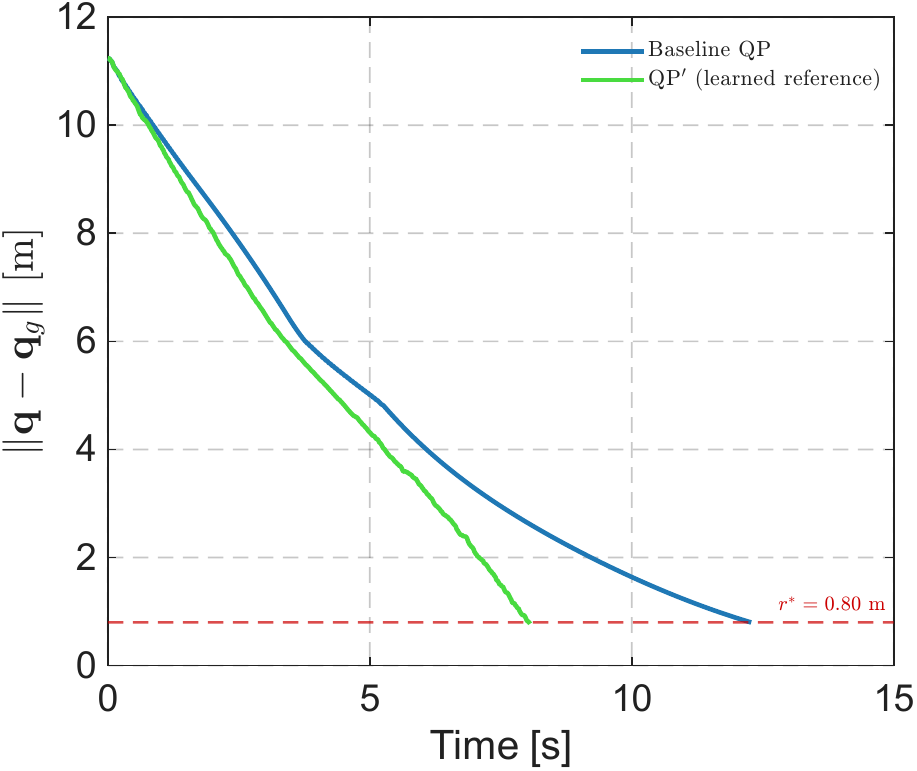}
  \caption{Goal distance: baseline vs.\ learned}
  \label{fig:rl_comparison}
\end{subfigure}
\caption{(a)~Near the
goal, the proposed controller settles at the marked
final position, while the sampled-data controller
($T=0.1\,\text{s}$) loops
repeatedly. (b)~Near closest approach, the sampled-data
controller makes a sharper, less direct avoidance manoeuvre.
(c)~In the corridor scenario, QP$'$ reaches $r^*=0.80\,
\text{m}$ roughly $4\,\text{s}$ sooner than the baseline,
consistent with the $36\%$ mean time-to-goal reduction over
$50$ trials.}
\vspace{-0.5cm}
\label{fig:traj_zoom_both}
\end{figure*}

\subsection{Main Stability Theorem}

\begin{thm}[Ultimate Bound on the Goal Distance]
\label{thm:stability}
Suppose Assumptions~\ref{ass:dist}--\ref{ass:init} hold and
QP~\eqref{eq:QP} is feasible along the closed-loop
trajectory under $\bu^*$ for all $t\geq0$. Let
$\bar\epsilon\triangleq\limsup_{t\to\infty}\epsilon^*(t)$ and
$r^*\triangleq2\delta\sqrt{k_y}/\eta$. Then:
(i) \textbf{Safety:} $h_i(\bx(t))\geq 0$ for
all $i$ and all $t\geq 0$ (Theorem~\ref{thm:safety});
(ii) \textbf{Ultimate bound:}
\begin{equation}
  \limsup_{t\to\infty}\rho(t) \;\leq\;
  r^*\sqrt{1+\frac{\eta\,\bar\epsilon}{2\delta^2k_y}}.
  \label{eq:UUB}
\end{equation}
In particular, $\limsup_{t\to\infty}\rho(t)\leq r^*$ whenever
$\epsilon^*(t)\to0$, with $r^*\to 0$ as $\delta\to 0$.
\end{thm}
\begin{proof}
Part~(i) is Theorem~\ref{thm:safety}. For part~(ii),
constraint~\eqref{eq:CLF} and Lemma~\ref{lem:DeltaV_bound}
give
\begin{equation}
  \dot V \leq -\eta V+\epsilon^*(t)+2\delta\sqrt{k_y}\sqrt V.
\end{equation}
By Young's inequality, $2\delta\sqrt{k_y}\sqrt V\leq
\frac{\eta}{2}V+\frac{2\delta^2k_y}{\eta}$, so
$\dot V\leq-\frac{\eta}{2}V+\frac{2\delta^2k_y}{\eta}
+\epsilon^*(t)$. By the comparison
lemma~\cite{khalil2002nonlinear},
\begin{equation}
  \limsup_{t\to\infty}V(t) \leq
  \frac{4\delta^2k_y}{\eta^2}+\frac{2\bar\epsilon}{\eta}
  =(r^*)^2\Bigl(1+\frac{\eta\,\bar\epsilon}{2\delta^2k_y}\Bigr),
\end{equation}
and $\rho(t)\leq\sqrt{V(t)}$ by Lemma~\ref{lem:rho_V}.
\end{proof}

\begin{remark}
\label{rem:r_star}
By~\eqref{eq:kkt_eps}, $\epsilon^*=\mu/(2p)$ is positive
whenever the CLF constraint is active, and it does not vanish
as $p\to\infty$ where the input cannot decrease $V$ (near
$e_x=0$, or while an RCBF opposes the CLF); this is why
\eqref{eq:UUB} keeps $\bar\epsilon$ explicit. The radius
$r^*=2\delta\sqrt{k_y}/\eta$ carries an explicit tradeoff:
raising $\eta$ shrinks $r^*$ but makes the CLF constraint
compete harder with the RCBFs during avoidance; raising
$k_y$ improves steering authority while enlarging $r^*$
only as $\sqrt{k_y}$; and $r^*$ scales linearly with
$\delta$. $\eta$ and $k_y$ are tuned jointly.
\end{remark}

\section{Numerical Simulation}\label{sec:sim}

All simulations use~\eqref{eq:unicycle} with
QP~\eqref{eq:QP} solved via \texttt{quadprog} (MATLAB), over
$T_\text{sim}=80\,\text{s}$ with $\Delta t=0.02\,\text{s}$.
The goal is $\bq_g=(8,8)\,\text{m}$; three obstacles at
$\bc_1=(3,4)$, $\bc_2=(5,6)$, $\bc_3=(7,3)\,\text{m}$ have
$r_i=0.6\,\text{m}$, $\rho_s=0.3\,\text{m}$, so
$\bar r_i=0.9\,\text{m}$. The disturbance
$\bd(t)=\delta[\cos(0.5t),\sin(0.5t)]^\top$ with
$\delta=0.1\,\text{m/s}$ attains $\norm{\bd(t)}=\delta$ at
every instant with continuously rotating direction, a
structure the controller does not exploit. The robot starts
at $(0,0)$, $\theta(0)=45^\circ$. Gains: $\eta=0.5$,
$\gamma=1.2$, $k_y=4.0$, $\lambda_\omega=0.3$, $p=200$,
$v_0=0.05\,\text{m/s}$, $\rho_\text{stop}=0.6\,\text{m}$,
$v_{\max}=3.0\,\text{m/s}$,
$\omega_{\max}=2.5\,\text{rad/s}$,
$h_\text{act}=8.0\,\text{m}^2$, giving
$r^*=2\delta\sqrt{k_y}/\eta=0.80\,\text{m}$.

\subsection{Results}\label{sec:results}
The trajectory bypasses all three obstacles smoothly,
slowing near the goal (Fig.~\ref{fig:results_traj}).
$\rho(t)$ falls from $\approx11.3\,\text{m}$ to within
$r^*=0.80\,\text{m}$ by $t\approx 30\,\text{s}$
(Fig.~\ref{fig:results_dist}), consistent with
Theorem~\ref{thm:stability}; a practical stopping rule
(holding $v=\omega=0$ once $\rho<\rho_\text{eq}\ll r^*$)
is applied thereafter. Distance to every
obstacle stays above $\bar r_i$ throughout
(Fig.~\ref{fig:results_cbf}), i.e.\ $h_i(\bx(t))\geq0$,
consistent with Theorem~\ref{thm:safety}.

\subsection{Monte Carlo Study}\label{sec:montecarlo}
Over $50$ trials with randomized phases and frequencies,
$\bd(t)=\delta[\cos(\omega_1t+\phi_1),
\sin(\omega_2t+\phi_2)]^\top$, robust tightening yields zero
violations; with tightening
disabled ($\sigma_i\equiv0$), violations occur in
$\approx30\%$ of trials, indicating that $\bar\Delta_i(\bx)$
is a practical necessity, not merely a theoretical safeguard.

\subsection{Learned Reference: Corridor Evaluation}
\label{sec:sim_rl}

To evaluate $\pi_\phi$, we use a scenario in which the direct
path to the goal threads the corridor between two obstacles:
$\bc_1=(5.99,4.01)$, $\bc_2=(4.01,5.99)\,\text{m}$
(separation $2.80\,\text{m}$, a $1.0\,\text{m}$ corridor),
so that both tightenings are active
($\sigma_1,\sigma_2>0$) along the corridor, with
$\bc_3=(7.50,2.00)\,\text{m}$ placed clear of $\bc_1$.

$\pi_\phi$ is a multilayer perceptron (MLP) with two hidden layers of $64$ units each and $\tanh$ activations. It is trained with PPO (clipping parameter $0.2$, discount factor $0.99$, generalized advantage estimation (GAE)~\cite{schulman2015high} with parameter $0.95$, and $3$ epochs per update) over $500$ episodes with randomized start pose, each capped at $40\,\text{s}$; other gains are as above, with $q_v=1$. Since a successful episode ends once
$\rho$ crosses $r^*$, $\pi_\phi$ rarely sees the final approach. The reference is therefore faded as
$(v_\text{RL},\omega_\text{RL})\leftarrow
\min\{1,\rho/(2r^*)\}\,(v_\text{RL},\omega_\text{RL})$, letting the CLF's terminal approach take over.

Over $50$ trials with randomized disturbance and start
jitter (Fig.~\ref{fig:rl_comparison}), both QP~\eqref{eq:QP}
and QP$'$ reach $r^*$ within $80\,\text{s}$ in $100\%$ of
trials with zero deadlocks; the learned reference reduces
mean time-to-goal from $13.3\,\text{s}$ to $8.5\,\text{s}$
($36\%$) and mean $\int\epsilon^*dt$ from $1.456$ to
$1.258$, indicating lower CLF--RCBF conflict along the
faster path.

\subsection{Learned Virtual Goal: Closed Gap}
\label{sec:sim_vg}

We place two obstacles symmetrically about the direct path,
$\bc_1=(5.67,4.33)$, $\bc_2=(4.33,5.67)\,\text{m}$, at separation
$D_{12}=1.90\,\text{m}$, with $\bc_3=(7.5,2.0)\,\text{m}$ and other
parameters as above. Here $\bar r_1+\bar r_2=1.80\,\text{m}$ and $D^\text{pass}_{12}=1.974\,\text{m}$: the gap is physically open ($0.70\,\text{m}$ between the obstacles) but closed for the robust controller by Corollary~\ref{cor:gap} (with the window~\eqref{eq:sigma_i}, $\sigma_i\geq0.998$ on the segment and $\beta_1+\beta_2\geq0.08$ there, so the conclusion is unchanged).
Whenever QP~\eqref{eq:QP}
is infeasible because of the bound $v\geq v_\text{min}(\rho)$, that
bound is relaxed to $v\geq-v_{\max}$ for the step, so that the
safety constraints remain enforced.

The policy $\varphi_\phi$ is a $2$-layer MLP ($64$-$64$, $\tanh$) with output scaled to $[-\tfrac{\pi}{2},\tfrac{\pi}{2}]$. It is queried every $0.5\,\text{s}$ and its output held in between, and it is trained with PPO (clip $0.2$, discount $0.99$ per decision, GAE parameter $0.95$, $3$ epochs per update, batch $1024$ decisions, learning rate $3\times10^{-4}$) for $1500$ episodes on reward~\eqref{eq:reward} with weights $(w_1,w_2,w_3,w_4)=(10,0.05,0.02,10^{-3})$, the effort term acting on $\varphi_\phi$, plus a bonus of $20$ on reaching the goal and a penalty of $0.05$ per step at which the speed bound is relaxed. Each episode draws a disturbance with random phases and frequencies, projected onto $\norm{\bd}\leq\delta$, and a start pose within $\pm1\,\text{m}$ and $\pm30^\circ$ of the nominal one, and is capped at $40\,\text{s}$.

\begin{figure}[t]
\centering
\includegraphics[width=0.75\linewidth]{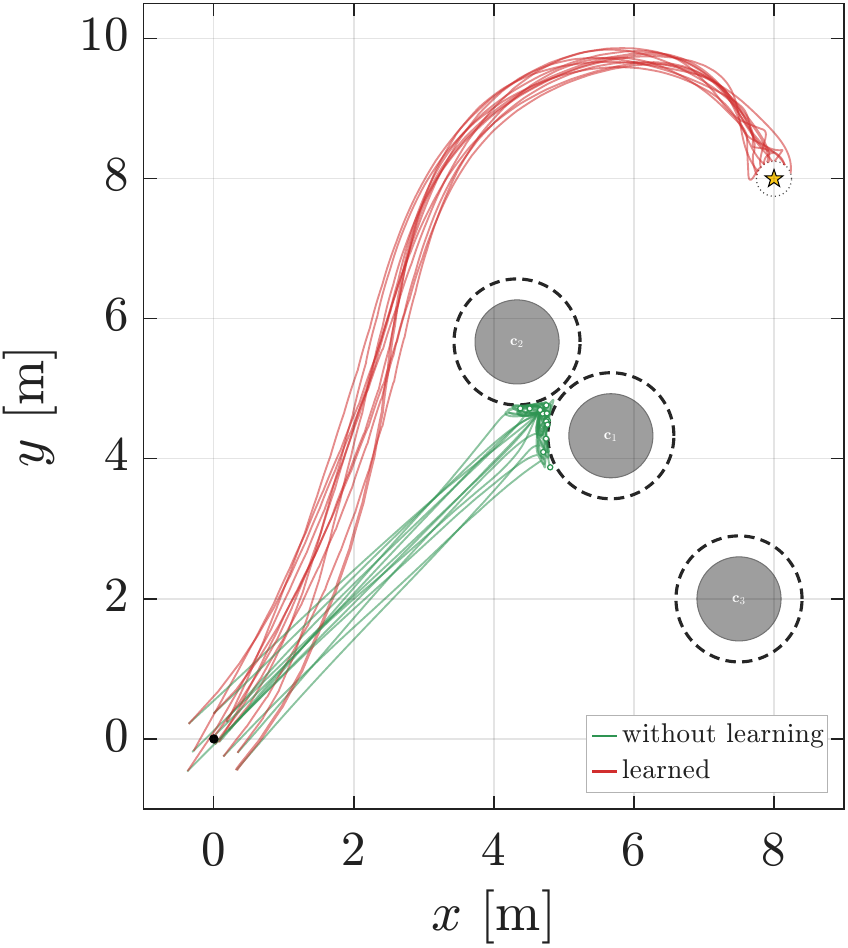}
\caption{Closed gap, $12$ of the $50$ test trials. Without learning (green) the robot stalls at the initiation of the gap, which is too narrow to cross under the robust constraints; with the learned virtual goal (red) it detours around the pair and comes to rest within $\rho_\text{eq}=0.25\,\text{m}$ of the goal.}
\label{fig:vg}
\vspace{-10pt}
\end{figure}

Table~\ref{tab:vg} reports $50$ test trials (start jitter $\pm0.5\,\text{m}$, $\pm15^\circ$, random disturbance). Without learning the robot drives to the initiation of the gap and stalls there in every trial, as Corollary~\ref{cor:gap} predicts, and the speed bound has to be relaxed at $58\%$ of the steps. With the learned virtual goal it detours around the pair and reaches the goal in all $50$ trials for both training seeds, with no relaxed step (Fig.~\ref{fig:vg}). Continuing each run under the stopping rule of Section~\ref{sec:results} with $\rho_\text{eq}=0.25\,\text{m}$, all $50$ trials of both seeds come to rest within $0.25\,\text{m}$ of the goal, at $t=20.6\pm2.3\,\text{s}$ and $20.7\pm2.3\,\text{s}$. No safety violation occurs in any of these runs ($h_i>0$ throughout), consistent with Proposition~\ref{prop:vg}; deadlock avoidance itself is an empirical result on this layout and carries no guarantee (Remark~\ref{rem:vg}).

\begin{table}[t]
\centering
\caption{Closed gap ($D_{12}=1.90\,\text{m}$): $50$ test trials per
row. Time to goal is mean $\pm$ std; ``relaxed'' is the fraction
of steps at which the lower speed bound was relaxed.}
\label{tab:vg}
\resizebox{\columnwidth}{!}{%
\begin{tabular}{@{}lccc@{}}
\toprule
CLF target & Reached & Time to goal (s) & Relaxed\\
\midrule
True goal (no learning) & 0/50 & -- & $58\%$\\
Learned virtual goal, seed 0 & 50/50 & $15.22\pm0.83$ & $0\%$\\
Learned virtual goal, seed 1 & 50/50 & $15.25\pm0.72$ & $0\%$\\
\bottomrule
\end{tabular}%
}
\end{table}

\subsection{Comparison With a Sampled-Data Robust CBF}
\label{sec:comparison}
Constraint~\eqref{eq:RCBF} is derived in continuous time and
re-solved every $\Delta t=0.02\,\text{s}$;
\cite{oruganti2023robust} instead holds a zero-order-hold
(ZOH) input over intervals of length $T$, requiring extra margin for
worst-case inter-sample drift. For a matched disturbance,
their tightening applied to \eqref{eq:hi} is
$\bar\Delta_i^\text{sd}(\bx)\triangleq\bar\Delta_i(\bx)
+T(\bar v+\bar\Delta_i(\bx))$, with
$\bar v=\sup\norm{f(\bx)+g(\bx)\bu}$. With
$T=0.1\,\text{s}$, $\bar v=v_{\max}$, the relative extra
margin $T(\bar v/\bar\Delta_i+1)$ is strictly decreasing in
$h_i$: the sampled-data controller is proportionally most
conservative near the obstacle, where conservatism is least
affordable. This shows up (Fig.~\ref{fig:traj_zoom_both}) as
slower convergence, looping near the goal, sharper avoidance
manoeuvres, and chattering in $v(t)$. Both controllers stay
safe throughout. The two controllers run at different update
rates, and the sampled-data margin buys an inter-sample
guarantee that the continuous-time constraint does not
provide; the difference is thus a structural cost of ZOH
control rather than a deficiency of either method.

\section{Conclusion}\label{sec:conclusion}
This paper presented a robust CLF-CBF-QP framework for safe
unicycle navigation among multiple obstacles under an unknown,
norm-bounded disturbance. A cross-track CLF built from body-frame
errors provides steering authority without a bearing angle. It is
paired with per-obstacle robust CBFs, each tightened by a closed-form disturbance bound. Whenever the QP is feasible, these
barriers guarantee forward invariance of every safe set, and the
goal distance is ultimately bounded by an explicit radius when the
slack vanishes. A learning-augmented variant, safe by construction, reduced the mean time-to-goal by $36\%$ in a corridor scenario without altering any guarantee. We also showed that a gap between two obstacles can be provably closed for the robust controller. For such a gap, a learned virtual goal for the CLF steered the robot around the obstacles while leaving every safety constraint untouched. Future work includes recursive-feasibility guarantees, moving obstacles via time-varying barrier functions, and hardware validation.

\bibliographystyle{IEEEtran}
\bibliography{References}

\end{document}